World Scientific
www.worldscientific.com

# Bouncing Universe with Exotic Radiation

C. Swastik

*Indian Institute of Astrophysics*
*Bangalore, 560034 India*
*swastikchowbay@iiap.res.in*

P. K. Suresh*

*School of Physics, University of Hyderabad*
*Hyderabad 500046, India*
*pkssp@uohyd.ernet.in*

Barun Maity

*National Centre for Radio Astrophysics (NCRA) — Tata*
*Institute of Fundamental Research (TIFR )*
*Ganeshkhind, Pune 411007, India*
*bmaity@ncra.tifr.res.in*

Scenario bouncing can give the cosmology singularity problem a possible way out. Finding a solution for the universe's bouncing model requires proper estimation of the state equation. We present two such state equations that give us the solution for a bounce. One such case is the exotic radiation, where we assumed that exotic radiation dominated the universe during the bounce occurrence. We also considered another case where quintom matter scenario existed previously, and the newly proposed exotic radiation scenario coexisted. In these two cases, we have shown that all the necessary conditions for the bounce are fulfilled. Such new ways certainly increase support and flexibility for the bouncing universe model.

## 1. Introduction

The bouncing model of the universe[1,2] is believed to provide solutions to the problem of singularity associated with the standard cosmological model. The standard model might account for most of the universe's observed features. Still, it encountered problems of horizon, flatness and singularity and defy solutions. The topic of singularity is the most difficult question in standard cosmology. The problem of singularity exists because of the classical nature of general relativity. Quantum gravity is therefore likely to be a viable scenario for solving the problem of singularity. Quantum gravity currently has many approaches but there is no conclusive model. Later, inflationary scenario was incorporated into mainstream cosmology to solve some of its problems but the problem of singularity remains[3–5] unresolved. In fact, the inflation situation itself has challenges of its own. It is therefore important to try and research alternate solution(s) in order to avoid singularity, rather than relying only on solutions to quantum gravity and inflation. Another such alternative approach is the bouncing universe model. Unlike the traditional cosmology paradigm, universe transits from a contracting process to an expanding bouncing scenario. In such a case, the scale factor at the bouncing point must be non-zero. Hence, the present universe can be considered as the result of a previous contracted phase with a non-zero minimal size to an expanding phase. In this way, it overcomes the question of singularity that is present in standard cosmology.

*Corresponding author.

Non-singular bouncing models received much attention among the various classes of bouncing models (see Refs. 6–8 for review). Besides having a non-zero scale factor at the bouncing point, there is yet another necessary condition for a bounce to take place. The state equation of the bounce candidate responsible must violate the null energy condition around the bouncing point[1] for a brief period of time. Satisfying this state is quite difficult for ordinary matter or radiation. These are therefore not viable candidates for ways to rebound. This situation forces us to hunt for appropriate alternatives for bouncing applications. There are actually multiple candidates for such solutions in non-single bouncing structures, and one such interesting candidate is the quintom matter.[9] The quintom model scenario is introduced in principle to understand the nature of the state equation for dynamic dark energy, and differs from the famous cosmological constant. It is shown that a bouncing solution with a phenomenological state equation is feasible for quintom matter. Studying the bouncing universe concept with quintom matter motivates us to investigate alternatives to bouncing universe with a newly proposed exotic radiation that is being incorporated in this paper. We designate the radiation suggested as exotic radiation, as its state equation deviates from the normal case of radiation. This paper mainly aims to study the possibility of bouncing solution with exotic radiation in non-singular bouncing model. We also study another scenario in which both the quintom matter and the proposed exotic radiation coexist in their equilibrium and find bouncing solutions in a non-singular bouncing model in their form of coexistence.

Note that the exotic radiation and quintom matter-exotic equilibrium radiation scenarios we find in this paper work are inspired by the quintom matter bouncing solution analysis. Therefore, the scale factor and state equation corresponding to the exotic radiation are introduced by similar approaches as the case of quintom matter. Although the state equation appears to differ slightly from that proposed by Cai *et al.*[2] in the case of exotic radiation, the scenarios are entirely different, making it unique from other proposed solutions. Comprehensive knowledge of the physical conditions at their equilibrium is necessary for understanding the type of scale factor in the quintom matter-exotic radiation equilibrium scenario. Since such specifics are insurmountable, and their estimation is beyond the scope of this paper, the forms of the scale factor and the equation of state of the equilibrium scenario case are taken as ansatz. This may be justified because it is possible to get a bouncing solution using these forms of the scale factor and the equation of state.

First, we discuss general features of bouncing model and conditions that are required to occur bounce very briefly. That is accompanied by a short discussion on the quintom matter's bouncing solution. Subsequent sections are respectively for the study of bouncing using the exotic radiation and quintom matter-exotic radiation equilibrium scenario. Finally, we present the discussions and conclusion of this paper.

## 2. Bouncing Solution with Quintom Matter

Einstein's general theory of relativity with the Friedmann–Lematire–Robertson–Walker (FLRW) metric can provide the necessary conditions to occur bounce. The Friedmann's equations for the FLRW metric with the perfect fluid as a source for gravity can be written by taking $c = 1$ and $8\pi G = 1$ as

$$\left[\frac{\dot{a}}{a}\right]^2 = \frac{\rho}{3} - \frac{k}{a^2}, \tag{1}$$

$$\left[\frac{\ddot{a}}{a}\right] = -\frac{1}{6}(\rho + 3p), \tag{2}$$

where $\rho$ and $p$, respectively, are the energy density and pressure for the perfect fluid. Here, $k$ is the curvature parameter and $a$ is the scale factor.

The Friedmann equation can be expressed in terms of the Hubble parameter ($H = \frac{\dot{a}}{a}$) and hence one can obtain

$$\dot{H} = \frac{k}{a^2} - \frac{1}{2}(\rho + p). \tag{3}$$

In order to obtain a bouncing solution, the Hubble rate which arises from the contracting phase with a negative value must increase as it is positive during the expansion phase. So, it is required that $\dot{H} > 0$. Thus, it implies that the violation of the null energy condition for a brief period of time around the bouncing point is essential to get a bouncing solution. In other words, the existence of bouncing solution can be understood in terms of the violation of the null energy condition.[1] Equation (3) for a flat universe can be written as

$$\dot{H} = -\frac{\rho}{2}(1 + \omega) > 0, \tag{4}$$

where the equation of state, $\omega = \frac{p}{\rho}$. The condition (4) requires that the universe must transit from its past with $\omega < -1$ to the later hot big bang phase with $\omega > -1$. This transition of the equation of state must be achieved to have a bouncing solution. Otherwise, when the universe enters into the later big bang phase after bounce period, the evolution of the universe continues with the same bouncing candidate carrying the equation of state $\omega < -1$. Consequently, it can lead to the big rip singularity. Hence, achieving the bounce and the required transition for the equation of state with ordinary matter or radiation or their equilibrium is insurmountable. Therefore, a candidate known as the quintom has been proposed to occur bounce with appropriate equation of state.

In the quintom matter scenario, the phenomenological equation of state is given by[2,10]

$$\omega(t) = -r - \frac{s}{t^2}, \tag{5}$$

for which the model parameters take value $r < 1$ and $s > 0$. The equation of state $\omega$ runs from negative infinity at $t = 0$ to the cosmological constant boundary at $t = \sqrt{\frac{s}{1-r}}$. It is shown that for the quintom matter with the equation of state (5), the Friedmann equation leads to

$$H(t) = \frac{2}{3} \frac{t}{(1-r)t^2 + s}, \tag{6}$$

and the scale factor in terms of the model parameters becomes

$$a(t) = \left[ t^2 + \frac{s}{1-r} \right]^{\frac{1}{3(1-r)}}. \tag{7}$$

Note that in absence of the model parameters, the Friedmann equation and the scale factor reduces to that of the ordinary matter dominated FLRW universe case. Further, it is shown that the bouncing solution exists in the quintom matter with the phenomenological equation of state.

The interesting study of quintom matter motivates us to explore for more such bouncing solution. We propose to have bouncing solutions with radiation of different types assuming a suitable equation of state. In such a case, the behavior of such radiation is expected to be different from that of the ordinary radiation. Therefore, in this paper, we use the nomenclature exotic radiation to denote it.

## 2.1. *Bouncing solution with exotic radiation*

We assume that the universe is dominated by exotic radiation. Following a similar approach of Eq. (5), we consider the equation of state to be

$$\omega(t) = \frac{1}{3} - \frac{4}{3} \left[ r + \frac{s}{t^2} \right], \tag{8}$$

where $s$ and $r$ are the model parameters which can take values $r < 1$ and $s > 0$. For the occurrence of a bounce, it is required that $\omega$ runs from negative infinity at $t = 0$ to the boundary at $t = \sqrt{\frac{s}{1-r}}$.

Then, for the above equation of state (8), the Friedmann equation leads to

$$H(t) = \frac{1}{2} \frac{t}{(1-r)t^2 + s}, \tag{9}$$

and hence the scale factor for exotic radiation is obtained as

$$a(t) = \left[ t^2 + \frac{s}{1-r} \right]^{\frac{1}{4(1-r)}}. \tag{10}$$

For simplicity, we normalize the scale factor $a = 1$ in the above equations at the bouncing point $t = 0$.

We study the behavior of the equation of state for the exotic radiation, the scale factor and the Hubble parameter. The obtained results are presented in Fig. 1. From the top panel of Fig. 1, it can be seen that the equation of state $\omega$ approaches to negative infinity at the bouncing point $t = 0$. We also see that the equation of state for exotic radiation can lead to the contracting phase for $t < 0$ to the subsequent expansion phase for $t > 0$. The middle panel of Fig. 1 shows that the scale factor for the exotic radiation is non-zero at the bouncing point $t = 0$. From the bottom panel of Fig. 1, we can observe that the Hubble parameter runs smoothly across the bouncing point at $t = 0$. These results show that the scale factor and the Hubble parameter satisfies the required conditions for a bouncing solution. Thus, it is possible to get a non-singular bouncing universe solution with the proposed exotic radiation. Note that the value of the model parameters, $r < 1$ and $s > 0$ holds strictly only for the bouncing case and in the absence of model parameters, the equation of state of the exotic radiation reduces to that of the standard radiation case for which bouncing solution does not exist.

## 2.2. *Bouncing solutions for equilibrium case*

In this section, we consider a scenario in which both the quintom matter and the exotic radiation coexist at their equilibrium and study whether bouncing solution exists for it or not. Since the actual form of the equation of state at the equilibrium scenario is again unknown, we assume the form of equation of state with a model parameter $m$ for the quintom-exotic radiation equilibrium. This is an ansatz which is motivated from the previous discussions. This EoS is given by

$$\omega = -1 + \frac{m}{3}\left[ 1 - r - \frac{s}{t^2} \right], \tag{11}$$

where $s$ and $r$ are once again the model parameters and, respectively, take value $r < 1$ and $s > 0$ for the bouncing scenario. In order to occur bounce, it is required that $\omega$ runs from negative infinity at $t = 0$ to the cosmological constant boundary at $t = \sqrt{\frac{s}{1-r}}$. The Hubble parameter corresponds to the scale factor for the equilibrium scenario is obtained as

$$H(t) = \frac{2}{m} \frac{t}{(1-r)t^2 + s}. \tag{12}$$

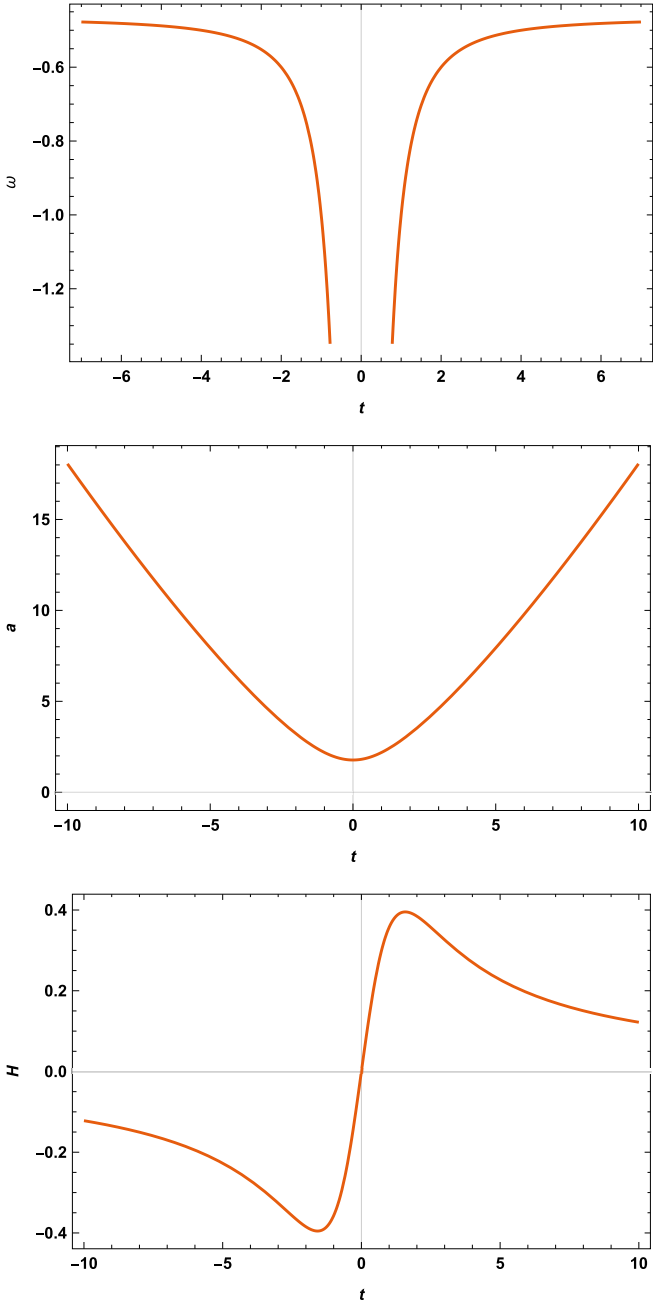

Fig. 1. Evolution of the equation of state, scale factor and the Hubble parameter for the exotic radiation case as a function cosmic time $t$ for $r = 0.6$ and $s = 1$.

Therefore, by using Eqs. (12) and (4), we obtain the scale factor for the quintom-exotic radiation equilibrium case as

$$a(t) = \left[t^2 + \frac{s}{1-r}\right]^{\frac{1}{m(1-r)}}, \tag{13}$$

In the absence of the model parameters $r$ and $s$, the scale factor varies as $t^{2/m}$, where the newly introduced parameter $m$ particularly takes only two extreme values 3 and 4 corresponding to the quintom matter and the exotic radiation, respectively. Hence, it reasonable to assume that the value of $m$ must lie between 3 to 4 in the equilibrium scenario case, but its actual value depends on the physical condition of equilibrium. The physical condition of the equilibrium scenario is not known. Hence, the exact computation of the value of $m$ at equilibrium is beyond the scope of this paper. However, one can estimate the value of $m$ for the equilibrium case.

Thus next, we need to study the behavior of the equation of state, the scale factor and the Hubble parameter for the coexistence case of the quintom matter and the exotic radiation at their equilibrium. The value of $m$ is unavoidable for this. Thus, the value of the parameter for the equilibrium case can be estimated as follows:

To estimate the value of $m$ for the equilibrium case, consider the Hubble parameter (12), hence

$$\frac{\dot{a}}{a} = \frac{2}{m} \frac{t}{(1-r)t^2 + s}. \tag{14}$$

Upon differentiating both sides of (12) with respect to $t$, we get,

$$\frac{\ddot{a}}{a} - \left(\frac{\dot{a}}{a}\right)^2 = \frac{2}{m} \frac{1}{(1-r)t^2 + s} - \frac{4t^2}{m} \frac{1}{\left(1-r\right)\left(t^2 + \frac{s}{1} - r\right)^2}. \tag{15}$$

Therefore, by using (14) in (15), we obtain

$$\frac{\ddot{a}}{a} = \frac{2}{m} \frac{1}{(1-r)t^2 + s} + \frac{4t^2}{\left(t^2 + \frac{s}{1-r}\right)^2} \left[\frac{1}{m^2(1-r)^2} - \frac{1}{m(1-r)}\right]. \tag{16}$$

Hence, by using the condition $\dot{H} > 0$ at the bouncing point that is at $t = 0$, we get

$$\frac{\ddot{a}}{a} = \frac{2}{ms} > 0. \tag{17}$$

Since $s > 0$, the condition (17) implies that $m$ must be positive. Thus, by interpolating $\frac{\ddot{a}}{a}$ for $s = 1$ with values of $m$ bounded between the quintom matter and the exotic-radiation, we obtained the value of $m$ for the equilibrium case as, $m \approx 3.42$. We obtained the value of $m$ by considering three points (contracting, expanding and bouncing) of which the values at two points are known (contracting and expanding). We performed linear interpolation to obtain the result. One can repeat the estimation of $m$ at the equilibrium case for the different allowed values of $s$ and it can be observed that its value remains invariant. Hence, the value of $m$ at the equilibrium case is independent of the model parameter as expected.

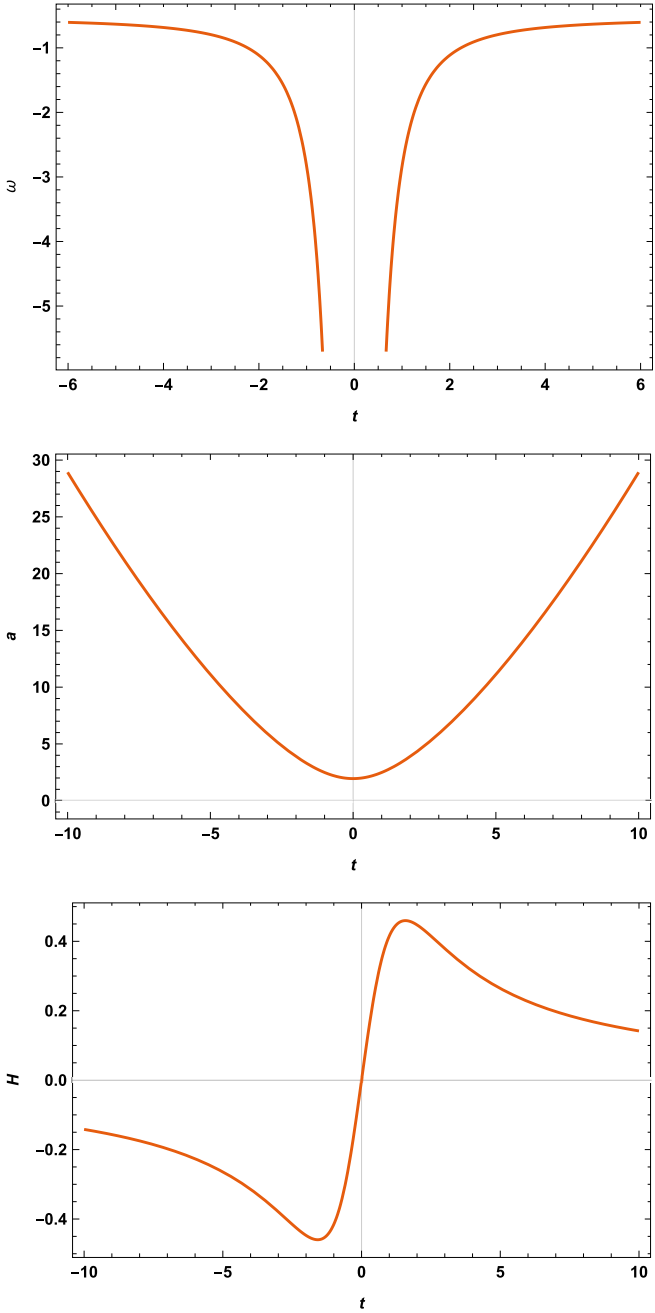

Fig. 2.   Evolution of the equation of state, scale factor and the Hubble parameter for the quintom matter-exotic radiation equilibrium case as a function cosmic time $t$ for $r = 0.6$, $s = 1$ and $m = 3.42$.

We study the scale factor, the Hubble parameter and state equation with the aforementioned estimated value of $m$, by normalizing the scale factor $a = 1$ at the bouncing point $t = 0$ for quintom matter and the exotic equilibrium case for radiation. The results obtained hereby be shown in Fig. 2. From the top panel of Fig. 2, it can be seen that at bouncing point $t = 0$ the state equation of the quintom matter exotic radiation equilibrium approaches negative infinity. The middle panel of Fig. 2 shows that the scale factor $a$ at the bouncing point $t = 0$ is non-zero at the bouncing point $t = 0$. From the bottom panel of Fig. 2, we can observe that the Hubble parameter runs smoothly through the bouncing point $t = 0$ for the equilibrium case of quintom matter-exotic radiation. This results show that the state equation and the Hubble function satisfy the conditions required for a bounce to take place. It can therefore be inferred that the coexistence of quintomic matter with exotic equilibrium radiation will lead to the evolution of the universe from the contracting phase of $t < 0$ to the corresponding expansion period of $t > 0$ with a bouncing point of $t = 0$ for the non-singular bouncing model type.

## 3. Discussion and Conclusion

The standard cosmology model based on the general relativity and cosmological principle is a very successful model which takes into account many observed features of the universe. However, it faces major problems such as flatness, horizon and singularity, etc. In order to resolve some of these problems, a scenario called inflation is introduced, but the problem of singularity remains a challenge. Meanwhile, a different kind of scenario is proposed to address the problem of singularity, known as the bouncing model. Under the bouncing cosmology model, the universe had a contracting phase accompanied by a bounce without the initial singularity. But for a bounce to occur, violation of the liable bouncing candidate's null energy condition is necessary. Since the ordinary matter and the radiation cannot fulfill the condition, it is difficult to achieve a bouncing solution with them. For non-singular bouncing models, therefore, multiple candidates are proposed to get a solution. An interesting candidate named quintom matter is proposed with a phenomenological state equation with the hope that the dynamic dark energy can be accounted for. It is shown that bouncing in a non-singular class of bouncing model is possible with quintom matter, which motivates us to implement a non-standard form of radiation with an appropriate state equation, called exotic radiation.

The main objective of this paper is to study the presence of bouncing solution in a non-singular form of bouncing cosmology of exotic radiation. As in the case of quintom matter, the equation of state corresponds to the exotic radiation is taken with two model parameters. It is found that the exotic radiation state equation can lead to the null energy condition being violated. A study of the associated scale factor, the Hubble parameter and the state equation is also found to satisfy the conditions required for bounce to occur. Furthermore, in the absence of the model parameters, the solution reduces to that of the standard case of radiation for which there is no bouncing solution.

The coexistence of quintom matter and the exotic radiation at their equilibrium is another possibility that is addressed in this paper. Therefore, it is investigated whether in non-singular form of bouncing cosmology, Â it can lead to a bouncing solution or not. Since the exact form of the scale factor is not known an additional parameter $m$ is assumed for a very general form. Consequently, without the real value of $m$, the corresponding Hubble parameter and its equation state are obtained. This is due to the fact that in order to obtain the actual value of the parameter $m$, detailed knowledge of the physical condition at equilibrium must be known. An estimate of $m$ is provided as $\approx 3.42$ however and is independent of the parameters of the model. Therefore, by examining the corresponding scale component, the Hubble function and the state equation, the presence of bouncing solution for the equilibrium case is investigated with the obtained value of $m$. The results obtained show that the coexistence of quintom matter in equilibrium with the exotic radiation satisfies the necessary condition for bounce to occur. Therefore, it can be concluded that for the coexisting phase of quintom matter, bouncing solution exists with the exotic radiation at their equilibrium.

In this paper, it is shown that bouncing solution exists in their equilibrium scenario with their corresponding equation of state for the exotic radiation situation as well as for the quintom matter and exotic radiation. The viability of bouncing cosmology with exotic radiation and the equilibrium of quintom matter-exotic radiation may be interesting, and may hopefully be examined in the future with appropriate observations.